\documentclass[12pt]{iopart}

\usepackage{iopams}
\usepackage{bm}
\usepackage{graphicx}
\usepackage{subfigure}

\newcommand{\td}{\rmd}
\newcommand{\te}{\rme}
\newcommand{\ti}{\rmi}

\newcommand{\msp}{\ \ }

\newcommand{\text}[1]{\textrm{\normalsize #1}}
\newcommand{\subscripttext}[1]{\textrm{\scriptsize #1}}
\newcommand{\onlinecite}[1]{\cite{#1}}
\newcommand{\eqref}[1]{\eref{#1}}

\newcommand{\lvert}{\vert}
\newcommand{\rvert}{\vert}

\newenvironment{align}{\begin{eqnarray}}{\end{eqnarray}}

\usepackage{psfrag}
\graphicspath{{./img/}{./}}

\usepackage{color}

\newcommand{\pgraph}[1]{}
\newcommand{\newstuff}[1]{{ #1}}

\begin{document}
\title[Nielsen--Olesen strings and vortex duality in 3+1 dimensions]{Condensing Nielsen--Olesen strings and the vortex--boson duality in 3+1 and higher dimensions}

\author{A.J. Beekman$^{1}$, D. Sadri$^{1,2}$ and J. Zaanen$^{1}$}

\address{${^1}$Instituut-Lorentz for Theoretical Physics, Universiteit Leiden, P.O. Box 9506, 2300 RA Leiden, The Netherlands}
\address{${^2}$Institute of Theoretical Physics, Ecole Polytechnique F\'ed\'erale de Lausanne (EPFL), CH-1015 Lausanne, Switzerland}
\ead{aron@lorentz.leidenuniv.nl}

\begin{abstract}
Dualities yield considerable insight in field theories by relating the weak coupling regime of one theory to the strong coupling regime of another. A prominent example is the ``vortex--boson'' (or ``Abelian-Higgs'',``$XY$'') duality in 2+1 dimensions demonstrating that the quantum disordered superfluid is equivalent to an ordered superconductor and the other way around. Such a duality structure should be ubiquitous but despite the simplicity of the complex scalar field theory  in 3+1 (and higher) dimensions a precise formulation of the duality is lacking. In 2+1 dimensions the construction rests on the fact that the topological excitations of the superfluid (vortices) are particle-like and the dual superconductor corresponds just with a conventional Bose condensate of vortices. Departing from the superfluid, the vortices in 3+1d are Nielsen--Olesen strings and the difficulty is in the construction of string field theory. We demonstrate that an earlier attempt \cite{Franz07} to construct the dual theory is subtly flawed. Relying on the understanding of the physics of the disordered superfluid in higher dimensions, as well as a gauge invariant formulation of the Higgs mechanism at work in this context, we derive the effective action for the dual string superconductor in 3+1d. This turns out to be a very simple affair: the string condensate just supports a massive compressional mode while it gives mass to the 2-form transversal photon that represents the remnant of the zero sound mode of the superfluid.  We conclude with the observation that the 2+1d superfluid--superconductor duality actually persists in all $D+1$ dimensions with $D \geq 2$: the condensates are formed from $D-2$-branes interacting
via $D-1$-form gauge fields but the form of the effective theory of the dual superconductor is eventually independent of dimensionality. \newstuff{
Finally, we demonstrate that Bose-Mott insulators support topological defects which are string-like in 3+1d. This surprising implication of duality may be seen in cold atom experiments.}
\end{abstract}

\submitto{\NJP}

\section{Introduction}
Dualities are among the most powerful weapons of field- and string theory. The Kramers--Wannier (weak--strong) dualities associated with theories controlled by Abelian symmetries are elementary examples. Among those the vortex (or ``Abelian-Higgs'' or ``$XY$'') duality in 2+1d is particularly famous \cite{Kleinert89a, FisherLee89, FradkinShenker79,HerbutTessanovic96, CvetkovicZaanen06a, NguyenSudbo99,HoveSudbo00,Zee00,Fisher04}. It states that the disordered, large coupling constant phase of the quantum $XY$ (global $U(1)$) system is equivalent to the small coupling constant  Higgs phase of an Abelian $U(1)$ superconductor interacting via a non-compact $U(1)$ gauge field. Since ``duality$^2  = 1$'', it is equally true  that the disordered Coulomb phase of this Higgs system is nothing else than the superfluid, the orderded phase of the global $U(1)$ theory.

\begin{figure}
\begin{centering}
\psfrag{t}{$t$}
\psfrag{x}{$x$}
\psfrag{y}{$y$}
 \includegraphics[width=8cm]{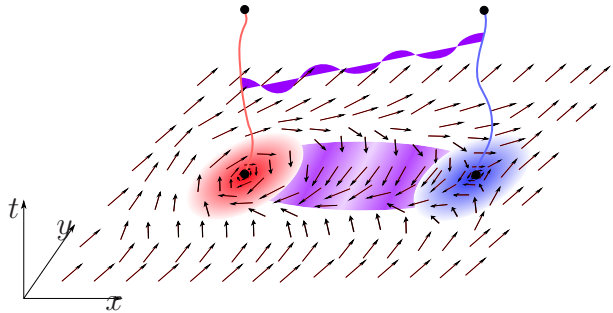}
\caption{Vortex (red) and anti-vortex (blue) interacting via a spin-wave fluctuation (purple) in a superfluid. \small{The vortices are defined completely in terms of the phase variable, which is frozen away from the defect pair, but wildly fluctuating in the neighbourhood of a vortex. Inside the core region, the arrows decrease in size to vanish at the origin, indicating that the phase in not well-defined. The vortices can be viewed as individual entities propagating in time; they interact through the exchange of a gauge particle, corresponding to an excited Goldstone mode.}}\label{fig:pointlike defects}
\end{centering}
\end{figure}

To set the stage, we will review in section \ref{sec:2+1d XY-model} the explicit derivation: the topological defects of the superfluid (vortices) are subjected to a long-range interaction that turns out to be identical to electrodynamics in  2+1 dimensions (see figure \ref{fig:pointlike defects}); vortices are particles in 2+1d and upon increasing the coupling constant the closed vortex--anti-vortex loops in spacetime expand such that eventually a `loop blowout' occurs at the quantum phase transition to the quantum disordered phase; this in turn corresponds with a tangle of free vortex worldlines that interact via $U(1)$ gauge bosons and this is nothing else than a superconductor/Higgs phase formed from the vortex condensate.

\begin{figure}
\begin{centering}
\psfrag{m}[][][.7]{$\mu$}
\psfrag{n}[][][.7]{$\nu$}
\psfrag{b}{$B_{\mu\nu}$}
\psfrag{t}{$t$}
\psfrag{x}{$x$}
\psfrag{y}{$y$}
 \includegraphics[width=8cm]{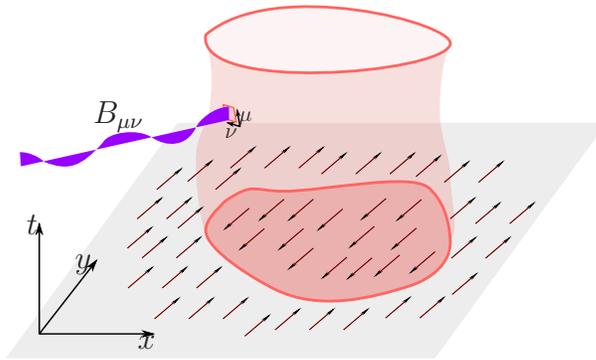}
\caption{A vortex worldsheet. \small{Cross section of a vortex loop in space that traces out a worldsheet. The third spatial dimension cannot be drawn. The phase $\varphi$ points away from or towards the vortex core. At each point in space the worldsheet is defined by a surface element with two spacetime indices $\mu$ and $\nu$, emitting a 2-form gauge field $B_{\mu\nu}$.}}\label{fig:vortex worldsheet}
\end{centering}
\end{figure}

Although such a duality should be perfectly general, its explicit construction is, even for a field theory as elementary as the complex scalar ($XY$) one, exclusively established in lower dimensions: we already alluded to the 2+1d case and of course the Kosterlitz--Thouless case in 1+1d is overly well known \cite{KosterlitzThouless70,HalperinNelson78,NelsonHalperin79,Young79}.  However, in 3+1 and higher dimensions these matters are not entirely settled. Increasing dimensionality renders the field theory to become simpler but another matter is to construct the duality. The problem is that the vortices turn in 3+1 dimensions into strings (``1-branes'', see figure \ref{fig:vortex worldsheet}), and in $D$+1 dimensions into $p= D-2$-branes using the string theory  convention where $p$ refers to the space dimensionality of the manifold. The disordered phase should then correspond with a `brane foam' taking the role of the vortex worldline tangle representing the Higgs condensate of the 2+1d case. Specifically for the 3+1d case the description of the `string condensate' involves knowledge of string field theory. Although vortices have a finite core size and are therefore strings of the Nielsen--Olesen variety \cite{NielsenOlesen73}---thereby much simpler than fundamental strings \cite{KakuKikkawa74,Witten86}---one encounters the difficulty that second quantization cannot be formulated for stringy matter. Accordingly, different from  matter formed from particles, an algorithm is lacking to compute the properties of such string condensates directly. The only example of a precise duality involving stringy topological excitations is the transversal field global Ising model in 2+1d \cite{Kogut79}. The strong coupling phase can be viewed as Bose condensate of Ising domain walls in space time \cite{Polyakov87}; remarkably, the Wegener duality \cite{Wegener71} demonstrates that this string condensate is actually the ordered (deconfining) phase of Ising
gauge theory, while the ordered Ising phase corresponds with the confining phase of the gauged theory.
 
As we will demonstrate in this paper, the string condensate  associated  with the dual of the global $U(1)$ superfluid in 3+1d is in fact quite similar to the Higgs condensate found in 2+1d, and we will argue that this is the case in all higher dimensions.  Much of the groundwork has already been done by Franz \cite{Franz07}, resting in turn on considerations regarding Nielsen--Olesen string field theory as developed in the string theory community in the 1970s and 1980s \cite{MarshallRamond75,Rey89}.  As reviewed in section \ref{sec:3+1d XY-model}, the stringy nature of the vortices implies that the long range vortex--vortex interactions are 
now encoded in Abelian 2-form gauge fields (figure \ref{fig:vortex worldsheet}). Deep in the strongly coupled disordered phase the amplitude fluctuations (`Higgs bosons') of the vortex string condensate can be ignored when the focus is on the effective theory describing the scaling limit. 

Franz and his predecessors \cite{MarshallRamond75,Rey89,Franz07} then rely on a seemingly obvious generalization of the Higgsing of particle condensates to construct the London-limit form of the effective action for the `stringy superconductor'.  We show that this Ansatz is actually incorrect.  In section \ref{sec:The Bose-Hubbard model} we review the Bose-Hubbard model which is a particularly convenient UV lattice regularization of the field theory. In this language the physical nature of the disordered superfluid becomes manifest: it is just a simple Mott insulator and we emphasize the emergent `stay at home' $U(1)$ gauge invariance that eventually controls the physics \cite{LeeNagaosaWen06}. The nature of the collective excitations in arbitrary dimensions becomes also obvious: this is just a doublet of massive `holon' and `doublon' excitations.
The problem with the minimal coupling construction of Franz  {\em et al.}  becomes then immediately obvious:
a vectorial phase is ascribed to the string condensate and this overcounts the number of massive photons (more precisely: photon polarizations) by one in 3+1 dimensions. More generally, in $D$+1 dimensions one would find $D$ photons while the real number of photons should be 2 in the Higgs phase regardless the dimensionality of the target space. This follows directly from the fact that one is dealing with an internal $U(1)$ symmetry. 

The understanding of string field theory just amounts to knowing the collective motions of the matter formed from the strings. By backward engineering from the answer (the Bose-Mott insulator) we show in section \ref{sec:3+1d XY-model} that the field theory associated with Nielsen--Olesen string condensate is embarrassingly simple: the ungauged  string superfluid just supports zero sound, a non-dissipative pressure wave as in the particle superfluid. The gauged (by 2-forms) string superconductor gives mass to the photons, and the condensate adds just a longitudinal photon like in the standard Higgs phase.  In section \ref{sec:2+1d XY-model} we show how matters can be understood in the 2+1d case in a language that avoids the artificiality of the redundant gauge degrees of freedom.

The key is that the vortices act as sources and sinks of supercurrents and therefore supercurrent is no longer conserved in the vortex condensate. One can write the dual action directly in terms of these supercurrents and in this way one sees immediately that the longitudinal photon is just the expression of the non-conservation of the supercurrent in the disordered phase. Formulated in this way the Higgs mechanism as of relevance to the duality becomes independent of dimensionality again and we use it to demonstrate that the dual string superconductor in 3+1d is gouverned by the same effective field theory as its 2+1d sibling. We conclude with the demonstration in section \ref{sec:Conclusions} that actually this wisdom holds in all higher dimensions, with  the perhaps surprising outcome  that the `$p$-brane' vortex condensates in high dimensions produce a long wavelength physics that is as simple as the dual superconductor in 2+1d.

Another result is that the Higgs phase supports topological defects of its own, like the Abrikosov vortices of type-II superconductors. These follow automatically in the duality construction, which we will show in section \ref{sec:Topological defects in the 3+1d Higgs phase}. But since the Higgs phase corresponds to a Bose-Mott insulator, this implies that a Mott insulator can also have stringlike vortices, which are induced by external superfluid order. We present an idea of how this could be seen in cold atom experiments.

We wish to stress that we are \emph{not} dualizing a vector gauge field coupled to complex scalar matter as the name ``Abelian-Higgs duality'' may suggest. Instead we are dualizing the scalar Goldstone mode of the superfluid; this literally corresponds to the Abelian-Higgs model only in 2+1 dimensions. Other works have considered dualizations involving two-form fields or string field theory \cite{Savit80,Orland83,Orland94,SeoSugamoto81,HoHosotani88}, but we point out that either their approach or physical motivation differ from ours. Also, in their original paper \cite{NielsenOlesen73} Nielsen and Olesen explicitly use the Abelian-Higgs model as one possible realization of finite core-size strings, and we feel therefore comfortable assigning their name to our vortices as well.

\section{Preliminary I: the Bose-Hubbard model}\label{sec:The Bose-Hubbard model}

The Bose-Hubbard model `at zero chemical potential' \cite{FisherLee89,FisherEtAl89} can be regarded as a convenient lattice regularization for the global $U(1)$ field theory we wish to consider. At present this model gets much attention since it is routinely produced in a literal fashion in cold bosonic atom systems living on an optical lattice \cite{GreinerEtAl02,BruderFazioSchoen05}. Let us shortly review this affair---all we need is that from the canonical formulation the physics can be directly read off regardless the dimension of the spacetime. 

We define the model on a hypercubic lattice employing conjugate boson creation and annihilation operators $b^\dagger_i$ and $b^{\phantom{\dagger}}_i$, with $[b^{\phantom{\dagger}}_i , b^\dagger_j] = \delta_{ij}$. The number operator is $n_i = b^\dagger_i b^{\phantom{\dagger}}_i$. The model is given by,
\begin{equation}\label{eq:Bose-Hubbard Hamiltonian}
H_{\text{BH}} = -\frac{t}{2}\sum_{\langle ij\rangle} (b^\dagger_i b^{\phantom{\dagger}}_j + b^\dagger_j b^{\phantom{\dagger}}_i) - \mu \sum_i n_i + U \sum_i (n_i-1)n_i.
\end{equation}

Here $t$ is the hopping or tunnelling parameter for neighbouring sites, $\mu$ the chemical potential and $U$ the on-site repulsion. We specialize to `zero chemical potential' (see e.g. \onlinecite{FisherLee89}) such that the average number of bosons per site is an integer.  Under this circumstance at some critical value of $U/t$ a transition will follow from the superfluid at small $U$  to the Mott-insulator at large $U$.  This corresponds with a literal realization of the lattice regularized quantum $XY$ model, with $U/t$ playing the role of  coupling constant.

The commutation relation for $n$ and $b$ is,
\begin{equation}\label{eq:commutation relation number and boson annihilation}
 [n_i , b^{\phantom{\dagger}}_j] = [b^\dagger_ib^{\phantom{\dagger}}_i, b^{\phantom{\dagger}}_j] = 0 + [b^\dagger_i, b^{\phantom{\dagger}}_j] b^{\phantom{\dagger}}_i = -\delta_{ij} b^{\phantom{\dagger}}_i.
\end{equation}
Similarly $[n_i , b^\dagger_j] = \delta_{ij}b^\dagger_i$. To recognize quantum phase dynamics consider the  substitution,
\begin{align}
 b^\dagger_i  &=& \sqrt{n_i} \te^{\ti \phi_i},  b^{\phantom{\dagger}}_i = \te^{-\ti \phi_i} \sqrt{n_i}.
\end{align}
Here $\phi_i$ is a real scalar variable. Using \eqref{eq:commutation relation number and boson annihilation}, the commutation relation for $n$ and $\phi$
follows,
\begin{align}
 [n_i , b^{\phantom{\dagger}}_j] & = \delta_{ij} b^{\phantom{\dagger}}_i  &\Rightarrow  [n_i ,\te^{-\ti \phi_j} \sqrt{n_j} ] = -\delta_{ij}\te^{-\ti \phi_j} \sqrt{n_j} \nonumber\\
 & &\phantom{mm}  [n_i ,\te^{-\ti \phi_j} ]= -\delta_{ij}\te^{-\ti \phi_j}.
\end{align}
This commutation relation corresponds to $[n_i,\phi_j ] = -\ti \delta_{ij}$, which can be checked via the Taylor expansion of the exponential. In this way we have switched from a description in terms of the conjugate variables $b$ and $b^\dagger$ into the conjugate variables $n$ and $\phi$. For the hopping term we find,
\begin{eqnarray}
 \frac{t}{2}\sum_{\langle ij\rangle} (b^\dagger_i b^{\phantom{\dagger}}_j + b^\dagger_j b^{\phantom{\dagger}}_i)  \to \\ 
\frac{t}{2}\sum_{\langle ij\rangle} (\sqrt{n_i} \te^{\ti(\phi_i - \phi_j)} \sqrt{n_j} + \sqrt{n_j} \te^{-\ti(\phi_i - \phi_j)} \sqrt{n_i} ).
\end{eqnarray}

We now regulate the filling by the chemical potential in such a way that there is a large integer number $n_0 \gg 1$ of bosons per site on average.  In this limit we can directly substitute for $\sqrt{n_i}$ the amplitude VEV $\sqrt{n_0};\ n_0 = \langle n_i \rangle$. The Hamiltonian \eqref{eq:Bose-Hubbard Hamiltonian} reduces after the amplitude condensation into the Hamiltonian describing phase dynamics,
\begin{equation}
 H = -t n_0 \sum_{\langle ij \rangle} \cos(\phi_i - \phi_j) + U \sum_i (n_i - 1)n_i.
\end{equation}
The chemical potential term is left implicit, being just responsible for the integer filling.  We recognize the quantum $XY$ model where the interaction term just codes for the rotor kinetic energy ($n_i$ is equivalent to the angular momentum operator of a $U(1)$ rotor). The continuum limit is obtained by naive coarse graining $\cos(\phi_{i+1} - \phi_i) \to \cos (\nabla \phi(x))$ and $n_i \to n(x)$, and by expanding the cosine,
\begin{equation}\label{eq:continuum BH model}
 H = - \int \td x \ \frac{1}{2}(\nabla \phi)^2 + \Omega \int \td x \ n(n-1),
\end{equation}
where we have rescaled the coefficients while $\phi$ is periodic, $\phi \rightarrow \phi +  2 \pi N$. After Legendre transformation the interaction term turns into the rotor kinetic energy in the Lagrangian ($n^2 \rightarrow \frac{1}{c^2}( \partial_{\tau} \phi )^2$), where $c$ is the speed of light resp. sound, and we obtain the effective phase action  for the compact $U(1)$ phase field $\varphi$,  being the point of departure of the duality constructions in the next sections, 
\begin{equation}\label{eq:spin wave}
 S_\subscripttext{superfluid} = \frac{1}{g}\int \td x\ \frac{1}{2} (\partial_\mu \varphi)^2,
\end{equation} 
where $g \sim \frac{U}{t} $ is the coupling constant.

This model has two stable fixed points, separated by a continuous phase transition governed by $XY$ universality in $D$+1 dimensions \cite{FisherEtAl89,Kleinert89a, Sachdev99, NguyenSudbo99, HoveSudbo00}. The scaling limit physics of the two stable states can be discerned by inspecting the $g\sim U/t \rightarrow 0$ (weak coupling) and $g \sim U/t \rightarrow \infty$ limits. In the weak coupling limit the $U(1)$ field breaks symmetry spontaneously and the theory describes the superfluid state. The small fluctuations in the phase field $\phi$ correspond either with a {\em single} Goldstone boson corresponding with the zero sound mode of the superfluid, or with the spin-wave of the quantum $XY$ model. The interpretation of the strong coupling limit departing from the lattice Bose-Hubbard model is perhaps less familiar. Consider a starting configuration with  the
integer number of bosons $n^0$ per site as imposed by the choice of chemical potential. The effect of the hopping will be to create a `doublon' $n^0 + 1$ and `holon' $n^0-1$ pair on two different sites $i$ and $j$: $n^0_i n^0_j \rightarrow (n^0 -1)_i (n^0+1)_j$. This will cost an energy $U$: the system turns into a 
Bose-Mott insulator. This in turn implies a phenomenon that is well-known in condensed matter physics \cite{AffleckMarston88,LeeNagaosaWen06}  but perhaps less so in  high energy physics. This simple Mott localization has in fact a profound consequence: it causes a `dynamical' emergence of a gauge symmetry. The global $U(1)$ symmetry controlling the weak coupling limit gets `spontaneously' gauged into a  compact $U(1)$ local symmetry. In the superfluid $b^{\dagger}_i \rightarrow \sqrt{n^0} e^{i\phi_i}$ and the phase $\phi_i$ is the global $U(1)$ of the superfluid. However, in the strongly coupled Mott insulator the number operator of the bosons is sharply quantized on every site,
\begin{equation}
\hat{n}_i | \Psi (\text{Mott}) \rangle   = n_0 | \Psi (\text{Mott}) \rangle
\label{Mott}
\end{equation}
and this in turn implies a gauge invariance,
\begin{eqnarray}
b^{\dagger}_i & \rightarrow & e^{i \alpha_i} b^{\dagger}_i \nonumber \\
b_i & \rightarrow & e^{- i \alpha_i} b_i \nonumber \\
\hat{n}_i & = & b^{\dagger}_i b_i \rightarrow \hat{n}_i.
\label{emerginggauge}
\end{eqnarray}
This is the celebrated `stay at home' $U(1)$ gauge invariance that has played a prominent role in the various
gauge theories for high-$T_c$ superconductivity developed for the fermionic incarnation of the Hubbard model \cite{LeeNagaosaWen06}.

One can also immediately read off the nature of the collective modes of the Bose-Mott insulator from the strong coupling limit.One can either remove or add a boson and the holon and doublon that are created can just freely delocalize on the lattice giving rise to massive excitations with a mass $\approx U/2$ given that the chemical potential is in the middle of the Mott gap. The continuum theory we are dealing with requires that the length scales are large compared to the lattice constant, a regime that is quite different from the lattice cut-off regime exposed here. The continuum description becomes literal close to the quantum phase transition but given adiabatic continuity we know that the strong coupling limits are still representative for the mode counting 
and so forth.  Starting close to the critical coupling on the  Mott side, the Mott physics takes over from the critical regime at the correlation length (or time).  At larger scales the `stay at home' gauge invariance takes over, although it now involves a volume  with a dimension set by the correlation length. Accordingly, one will find the pair of degenerate propagating holon/doublon modes that appear as bound states that are pulled out of the critical continuum \cite{CvetkovicZaanen06a}. Similarly one finds on the superfluid side of the quantum critical point the single zero sound Goldstone boson at energies less than the scale set by the renormalized superfluid stiffness that disappears at the quantum critical point.

The simple features we have discussed in this section are generic and completely independent of the dimensionality of spacetime. Although perhaps unfamiliar, they are easily identified in the context of the standard vortex duality in 2+1d as discussed in the next section. In turn, they will be quite helpful in giving a firm hold  in our construction of the duality in higher dimensions.

\section{Preliminary II: Duality in the 2+1d $XY$-model}\label{sec:2+1d XY-model}

Let us now review the very well known vortex duality in 2+1 dimensions. This section is largely intended as a template for the development of the duality in 3+1d but towards the end of this section we do discuss a non-standard way of interpreting the dual superconductor, focussing on the physical currents and their conservation laws, thereby avoiding the `auxiliary' gauge fields of the standard duality. We also demonstrate 
how the physical emergent `stay at home' gauge principle of the Mott insulator  arises in the dual superconductor framework. These motives are important for decyphering the duality in higher dimensions. 

The first step in the 2+1d duality is to establish that vortices are just like charged particles in 2+1d electrodynamics. The quantum partition sum associated with the action \eqref{eq:spin wave} is,
\begin{equation}\label{eq:generating functional}
 Z = \int \mathcal{D} \varphi\ \te^{\ti \int \mathcal{L}} = \int \mathcal{D} \varphi\ \te^{\ti \int \frac{1}{2g} (\partial_\mu \varphi)^2}.
\end{equation}
turning into,
\begin{equation}\label{eq:dual generating functional}
 Z_\subscripttext{dual} = \int \mathcal{D} \varphi \mathcal{D}\xi_\mu\ \te^{\ti \int \frac{1}{2}g \xi_\mu\xi_\mu + \ti \xi_\mu \partial_\mu \varphi},
\end{equation}
by the Hubbard--Stratonovich transformation. The auxiliary $\xi_\mu$ field are dual variables; in canonical language going from $\varphi$ to $\xi_\mu$ amounts to a  Legendre transform; the dual variables are in fact the canonical momenta $\xi_\mu = \frac{\partial \mathcal{L}}{\partial(\partial_\mu \varphi)}$. These are also the Noether currents related to the tranformation $\varphi(x) \to \varphi(x) + \alpha$ under which \eqref{eq:generating functional} is invariant. When vortices are present in the superfluid, the otherwise smooth phase variable $\varphi$ is singular inside the core region (see figure \ref{fig:pointlike defects}). We therefore split it into a smooth and a multi-valued part: $\varphi = \varphi_\subscripttext{smooth} + \varphi_\subscripttext{MV}$. The multi-valued part denotes vortices of winding number $N$ via,
\begin{equation}\label{eq:winding number}
 \oint \td \varphi_\subscripttext{MV} = 2 \pi N.
\end{equation}
The smooth fields are integrated by parts,
\begin{equation}
 Z_\subscripttext{dual} = \int \mathcal{D} \varphi_\subscripttext{MV} \mathcal{D} \varphi_\subscripttext{smooth} \mathcal{D}\xi_\mu\ \te^{\ti \int \frac{1}{2}g \xi_\mu\xi_\mu + \ti \xi_\mu \partial_\mu \varphi_\subscripttext{MV} - \ti \varphi_\subscripttext{smooth}(\partial_\mu \xi_\mu)}
\end{equation}
and $\varphi_\subscripttext{smooth}$  is as a Lagrange multiplier that after integration yields the constraint $\partial_\mu \xi_\mu =0$. We recognize that the $\xi_{\mu}$ fields are just coding for the space- and time components of the supercurrent. The constraint is just the continuity equation expressing that supercurrents are conserved in the superfluid as long as the phase field is single-valued. In 2+1d  this continuity can be imposed  by expressing the current as the curl of non-compact $U(1)$ 1-form gauge field $A_{\mu}$,
\begin{equation}\label{eq:superflow to gauge field}
\xi_\mu(x) = \epsilon_{\mu\nu\lambda} \partial_\nu A_\lambda(x).
\end{equation}
such that $\xi_\mu$ is invariant under gauge transformations $A_\lambda \to A_\lambda + \partial_\lambda \varepsilon$ for any real scalar field $\varepsilon(x)$. The path integral over $\xi_\mu$ can be replaced by one over $A_\lambda$ provided one divides out the gauge volume which we leave implicit. We apply this substitution and perform another integration by parts to obtain,
\begin{equation}\label{eq:dual gauge action}
 Z_\subscripttext{dual} = \int \mathcal{D} \varphi_\subscripttext{MV} \mathcal{D}A_\lambda\ \te^{\ti \int \frac{1}{2} g(\epsilon_{\mu\nu\lambda} \partial_\nu A_\lambda)^2 + \ti A_\mu  J^V_{\mu}},
\end{equation}
where we define  $J^V_\lambda = \epsilon_{\lambda \mu \nu} \partial_\mu\partial_\nu \varphi_\subscripttext{MV}$. Because $\varphi_\subscripttext{MV}$ is multi-valued, the derivatives do not commute (cf. \eqref{eq:winding number}). These are the vortex currents associated with the multi-valued field configurations. On the one hand this expresses the fact that vortices act as sources and sinks of the supercurrents such that the latter are no longer conserved in the presence of vortices. At the same time, the simple derivation in the above demonstrates that the physics of the $XY$ model in 2+1 dimensions is indistinguishable of electromagnetism (EM) with the vortices taking the role of electrically charged particles that interact via photons that are the `force representatives' of the Goldstone bosons of the superfluid.

\begin{figure}
\begin{centering}
\psfrag{t}{$t$}
\psfrag{x}{$x$}
\psfrag{y}{$y$}
\psfrag{u}{$\tau$}
\psfrag{L}{$L$}
\psfrag{T}{$T$}
\psfrag{a}{$\parallel$}
\psfrag{e}{$\perp$}
\psfrag{p}{$p_\mu$}
 \includegraphics[width=6cm]{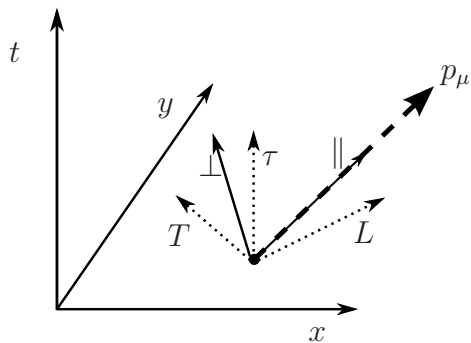}
\caption{Coordinate systems \small{We often use two coordinate systems related to the momentum $p_\mu$ of the gauge particle. In the $(\tau, L, T)$-system (dotted lines), the temporal direction is preserved, and the spatial ones are separated in longitudinal and transveral. This system is useful in the Coulomb gauge and when Lorentz invariance is broken. In a relativistic context, more useful is the $(\parallel, \perp, T)$-system (solid lines), where the $\tau$ and $L$-directions are rotated so that one is parallel to the spacetime momentum $p_\mu$. This direction $\parallel$ is also called longitudinal. The spatial-transversal directions are the same as in the previous system. In higher dimensions, there are simply more spatial-transversal directions.}}\label{fig:coordinate systems}
\end{centering}
\end{figure}

As long as the vortices are static, or when they are locked up in closed loops of vortex--anti-vortex pairs, the superfluid order is preserved and this represents the Coulomb phase in the EM dual. The vortex--vortex interactions have both static (Coulomb force) and dynamic (propagating photon) components. We adopt a coordinate system in Fourier space (figure \ref{fig:coordinate systems})  with temporal, longitudinal and transversal directions $( \tau , L , T)$ relative to the momentum $\ti \partial_\mu \to p_\mu  = (\omega ,q ,0)$. In these coordinates the Coulomb gauge $\nabla \cdot \mathbf{A} = 0$ turns into the requirement  $q A_L = 0$. In this gauge the Lagrangian takes the simple form,
\begin{equation}\label{eq:Coulomb gauge}
 \mathcal{L}_\subscripttext{Coulomb gauge} = \frac{1}{2}gq^2 A_\tau A_\tau + \frac{1}{2}gp^2 A_T A_T + \ti A_\tau J_\tau +\ti A_T J_T. 
\end{equation}
We see that the vortex sources emit gauge fields with propagators,
\begin{align}
 \langle \langle  A_\tau(p) A_\tau(0) \rangle \rangle &= \frac{1}{gq^2},\\
\langle \langle  A_T(p) A_T(0) \rangle \rangle &= \frac{1}{g(\omega^2 + q^2)} = \frac{1}{gp^2}.
\end{align}
We recover  the static long-range Coulomb force with a $\frac{1}{\lvert \mathbf{r} \rvert}$-potential, and the single, transversely polarized massless propagating photon of 2+1d EM, respectively. The static `photon' reflects the well known fact that static vortices in 2d interact via a Coulomb potential, and the transversal photon is just zero sound while in the dual `force' language it becomes explicit that this Goldstone boson can propagate forces between sources and sinks of supercurrent.  We stress that this correspondence between the `$XY$ universe' and 2+1d EM with scalar matter is quite accidental for the 2+1d case. We will see in the next section that this correspondence is completely lost in higher dimensions.

Upon increasing the coupling constant the vacuum will be populated by an increasing density of closed vortex--anti-vortex loops that grow in size. The quantum phase transition to the quantum disordered/Mott insulating phase occurs when the `loops blow out':  when the coupling constant is large enough that  the typical length of the vortex worldlines becomes of order of the system size, destroying the superfluid order.  The tangle of (anti-)vortex worldlines that forms  is like a tangle of charged particle worldlines in spacetime and this just corresponds with a relativistic superconductor/Higgs condensate \cite{CvetkovicZaanen06a,FossheimSudbo04,Kleinert89a}. This vortex condensate is described by  a complex scalar order parameter field $\Psi(x) = \lvert \Psi(x) \rvert \te^{\ti \chi(x)}$ with the currents associated with the vortex condensate,
\begin{equation}\label{eq:Higgs vortex current}
 J^V_\lambda = \ti\big((\partial_\lambda \overline{\Psi}) \Psi - \overline{\Psi} \partial_\lambda \Psi\big),
\end{equation}
while the order parameter $\Psi$ is gouverned by a Ginzburg--Landau action,
\begin{equation}
\mathcal{L}_\subscripttext{condensate}  = \frac{1}{2} \lvert D_\mu \Psi\rvert^2  + \frac{1}{2} m^2 \lvert\Psi\rvert^2 + \frac{1}{4} \omega\lvert\Psi\rvert^4 - \frac{1}{4}gF_{\mu\nu}F^{\mu\nu}.
\end{equation}
This can be explicitly derived using statistical physics methods, see references mentioned. Across the phase transition the parameter $m^2$ becomes negative, and the action is minimal at $\lvert\Psi(x)\rvert = \sqrt{\frac{-m^2}{g}} \equiv \Psi_0$. Only the condensate phase $\chi$ remains as a degree of freedom. The vortex condensate interacts with the `$XY$' gauge fields $A_{\mu}$ in the same way as a electromagnetically charged Bose condensate and therefore its  order parameter  is minimally coupled to the gauge field,
\begin{equation}\label{eq:minimal coupling}
\lvert \partial_\mu \Psi\rvert^2 \to \lvert D_\mu \Psi\rvert^2 = \lvert (\partial_\mu - \ti A_\mu) \Psi\rvert^2 = \Psi_0^2 (\partial_\mu \chi - A_\mu)^2.
\end{equation}
Referring to \eqref{eq:Higgs vortex current}, it indeed contains the coupling $\ti A_\lambda J^V_\lambda \to A_\lambda (\partial_\lambda \overline{\Psi}) \Psi + \text{h.c.}$. We have now a full view on the 2+1d vortex duality: the quantum disordered superfluid is from the dual perspective identical to the ordered superconductor. 

Since dual$^2 = 1$ it is equally true that the quantum disordered superconductor (the Coulomb phase of the gauge theory) can be  viewed as the ordered superfluid. \newstuff{This is done in a very similar way:

We linearize the  coupling term via an auxiliary field $v_\mu$ (constant terms are suppressed),
\begin{equation}\label{eq:2+1d dual squared}
 \mathcal{L} = \frac{1}{2} \frac{1}{\Psi_0^2 } v_\mu^2 + \ti v_\mu (\partial_\mu \chi - A_\mu) + \frac{1}{2} g(\epsilon_{\mu\nu\lambda} \partial_\nu A_\lambda)^2.
\end{equation}
The variable $\chi(x)$ describes the phase of the condensate field $\Psi$. Dual (Abrikosov) vortices are singularities in this phase field, and therefore we split it into a smooth and a multi-valued part: $\chi = \chi_\subscripttext{smooth} + \chi_\subscripttext{MV}$. On the smooth part, we can perform integration by parts and then integrate it out as a Lagrange multiplier for the condition $\partial_\mu v_\mu= 0$. This condition can be explicitly enforced by writing $v_\mu$ as the curl of another  gauge field: $v_\mu = \epsilon_{\mu\nu\lambda} \partial_\nu Z_\lambda$. This gives, after rescaling $A_\lambda \to \frac{1}{\sqrt{g}} A_\lambda$,
\begin{equation}
 \fl \mathcal{L} = \frac{1}{2} \frac{1}{\Psi_0^2 } (\epsilon_{\mu\nu\lambda} \partial_\nu Z_\lambda)^2 +\frac{1}{2} (\epsilon_{\mu\nu\lambda} \partial_\nu A_\lambda)^2 +\ti \epsilon_{\mu\nu\lambda} \partial_\nu Z_\lambda\partial_\mu \chi_\subscripttext{MV} +\frac{1}{\sqrt{g}} A_\mu\epsilon_{\mu\nu\lambda} \partial_\nu Z_\lambda  .
\end{equation}
On each of the last two terms we can perform integration by parts. The first of these is then the coupling of the gauge field $Z_\lambda$ to the Abrikosov vortex current $K_\lambda = \epsilon_{\lambda\mu\nu} \partial_\mu \partial_\nu \chi_\subscripttext{MV}$. Furthermore we see that the gauge field $A_\lambda$  only shows up in the combination $\xi_\mu = \epsilon_{\mu\nu\lambda} \partial_\nu A_\lambda$. We can now integrate out $\xi_\mu$ to leave a Meissner term for the gauge field $Z_\lambda$,
\begin{equation}
 \mathcal{L} = \frac{1}{2} \frac{1}{\Psi_0^2 } (\epsilon_{\mu\nu\lambda} \partial_\nu Z_\lambda)^2 + \frac{1}{2g}Z_\lambda^2 + \ti Z_\lambda K_\lambda.
\end{equation}
The interpretation of this action is as follows: the $XY$-disordered (Higgs/Meissner) phase is a state where Abrikosov vortices $K_\lambda$ source gauge fields $Z_\lambda$ that mediate interactions between those vortices. These interactions are however short-ranged due to the mass term for $Z_\lambda$.

Now we envisage that the Abrikosov vortices proliferate. They must then be described by a collective field $\Phi$ just as we did for the superfluid vortices  in \eqref{eq:Higgs vortex current}. The full Lagrangian reads, after rescaling $Z_\lambda \to \Psi_0 Z_\lambda$,
\begin{equation}
 \fl \mathcal{L} = \frac{1}{2} (\epsilon_{\mu\nu\lambda} \partial_\nu Z_\lambda)^2 + \frac{\Psi_0^2}{2g}Z_\lambda^2+ \frac{1}{2}\lvert(\partial_\mu - \ti \Psi_0 Z_\mu)\Phi\rvert^2 + \frac{1}{2} M^2 \lvert\Phi\rvert^2 + \frac{1}{4} W \lvert \Phi \rvert^4.
\end{equation}
We see that the disorder parameter $\Psi_0$ acts as a charge for the coupling of the gauge field $Z_\mu$ to the Abrikosov vortex field $\Phi$. When the Abrikosov vortices proliferate, they destroy the dual superconducting order, implying that $\Psi_0 \to 0$. The vortex field $\Phi$ then decouples from the gauge field $Z_\mu$, and we are left with the Landau action for a neutral superfluid:
\begin{equation}\label{eq:neutral superfluid}
 \mathcal{L} =  \frac{1}{2}\lvert\partial_\mu\Phi\rvert^2 + \frac{1}{2} M^2 \lvert\Phi\rvert^2 + \frac{1}{4} W \lvert \Phi \rvert^4.
\end{equation}
Indeed, through another duality construction we are back to our starting point of superfluid order. Which side is the `original' and which the `dual' one is completely up to one's own interpretation.
}

How to count the modes of the superconductor? It is the standard relativistic Abelian-Higgs affair. 
Choose coordinates $( \parallel , \perp, T )$ with $\parallel$ parallel to the spacetime momentum $p_\mu$, and $\perp$ perpendicular to both $\parallel$ and $T$ (figure \ref{fig:coordinate systems}). In this system the momentum becomes $p_\mu = (p , 0 ,0)$. We see that the Higgs phase $\chi$ couples only to the parallel direction,
\begin{align}\label{eq:2+1d Higgs action}
\mathcal{L}_\subscripttext{dual Higgs} &= -\frac{1}{2}g(\epsilon_{\mu\nu\lambda} \partial_\nu A_\lambda)^2 + \frac{1}{2}\lvert (\partial_\mu - \ti A_\mu) \Psi \rvert^2 \nonumber\\
&\to   \frac{1}{2}(p^2 + \Psi_0^2)(A_\perp^2 + A_T^2) +  \frac{1}{2} \Psi_0^2(p \chi - A_\parallel)^2 .
\end{align}
This action is invariant under the combined gauge transformations $A_\parallel \to A_\parallel + p \varepsilon$ and $\chi \to \chi + \varepsilon$. One possible gauge fix is the unitary gauge $\chi \equiv 0$ and in this way one shuffles the condensate mode into the ``longitudinal photon'' $A_\parallel$. Alternatively, we can choose the Lorenz gauge $pA_\parallel \equiv 0$, in which this degree of freedom is indeed seen to originate in the condensate field $\chi$. The field $A_\perp$ corresponds to the now short-ranged Coulomb force, and $A_T$ and $A_\parallel$ form a degenerate pair of  massive propagating modes. This matches precisely the expectations that follow from the Bose-Hubbard model; in the superfluid/Coulomb phase a single massless propagating mode is present corresponding with the phase mode/photon. In the dual superconductor one finds a pair of massive propagating modes corresponding with the Higgsed transversal- and longitudinal photons: these correspond with the holon and doublon excitations  of the Bose-Mott insulator while the Higgs mass of the dual superconductor just codes for the Mott gap---see  \onlinecite{CvetkovicZaanen06a} for further details.

Up to this point we have just reviewed the standard 2+1d vortex duality. For the purpose of understanding of how the duality works in higher dimensions we now want to discuss the duality from a  different viewpoint that is in a way more general and flexible. The culprit in the above is the emphasis on the gauge fields $A_{\mu}$. In fact, these are introduced as just a convenient trick  to impose  the continuity equation associated with the supercurrents of the superfluid in the absence of vortices. In fact, one can avoid the gauge fields entirely in the construction of the duality, and equally well in the description of the Higgs phase, by just formulating matters in terms of  the physical currents $\xi_{\mu}$. In a first step, by just formally integrating out the condensate phase field $\chi$ in the condensed superconductor, and using \eqref{eq:superflow to gauge field} to re-express the gauge fields back in the physical supercurrents, the effective action \eqref{eq:2+1d Higgs action} can be written as,
\begin{equation}\label{eq:Higgs superflow}
 \mathcal{L}_\subscripttext{Higgs, superflow} = \frac{1}{2}g\xi_\mu^2 + \frac{1}{2}\xi_\mu \frac{\Psi_0^2}{-\partial^2} \xi_\mu, 
\end{equation}
where the first term is just the action of the superfluid while the second `gauge invariant' Higgs term demonstrates that the supercurrents have now only short-range correlations, since they are no longer conserved in the presence of the vortex condensate. However, the latter statement also implies that we have to drop the continuity equation associated with the currents of the superfluid and we can no longer parametrize these currents by gauge fields! The fact that $\partial_{\mu} \xi_{\mu} \neq 0$ implies that the $\xi_{\mu}$ fields now also contain longitudinal components. We can now use the general wisdom of the Helmholtz decomposition, stating  that  a sufficiently smooth vector field $\xi_\mu$ is the sum of an irrotational (curl-free) and a solenoidal (divergence-free) part,
\begin{equation}\label{eq:Helmholtz decomposition}
 \xi_\mu = \partial_\mu \psi + \epsilon_{\mu\nu\lambda} \partial_\nu A_\lambda.
\end{equation}
When current is conserved $\partial_\mu \xi_\mu =0$, one sees that the irrotational part is restriced $\partial^2 \psi = 0 \Rightarrow \psi = 0\ \forall p \neq 0$. But in the Higgs phase, the constraint is released and the additional component shows up. From the decomposition it is clear that the two parts are orthogonal, so that,
\begin{equation}
 \xi_\mu^2 = (\partial_\mu \psi)^2 + (\epsilon_{\mu\nu\lambda} \partial_\nu A_\lambda)^2.
\end{equation}
and by inserting this in \eqref{eq:Higgs superflow} we find an effective action,
\begin{align}\label{eq:Higgs superflow1}
 \mathcal{L}_\subscripttext{Higgs, superflow} &= \frac{1}{2}g\xi_\mu^2 + \frac{1}{2}\xi_\mu \frac{\Psi_0^2}{-\partial^2} \xi_\mu \nonumber\\
&= \frac{1}{2}(p^2 + \frac{\Psi_0^2}{g}) \psi^2 +\frac{1}{2}(p^2 + \frac{\Psi_0^2}{g})(A_\perp^2 + A_T^2),
\end{align}
where we have rescaled $\xi_\mu \to \frac{1}{\sqrt{g}} \xi_\mu$ in the second line. This describes correctly the degenerate pair of massive `photons' ($\psi$ and $A_T$) that actually code for the holon-doublon excitations of the Mott insulator, supplemented with the Coulomb force $A_\perp$.

\begin{figure}
 \begin{center}
\hspace{.2cm}
$\frac{1}{\sqrt{2}}\Bigg( \Bigg\lvert $
\psfrag{A}{$A$}
\psfrag{B}{$B$}
\psfrag{r}{$r$}
\raisebox{-1cm}{\includegraphics[height=2cm]{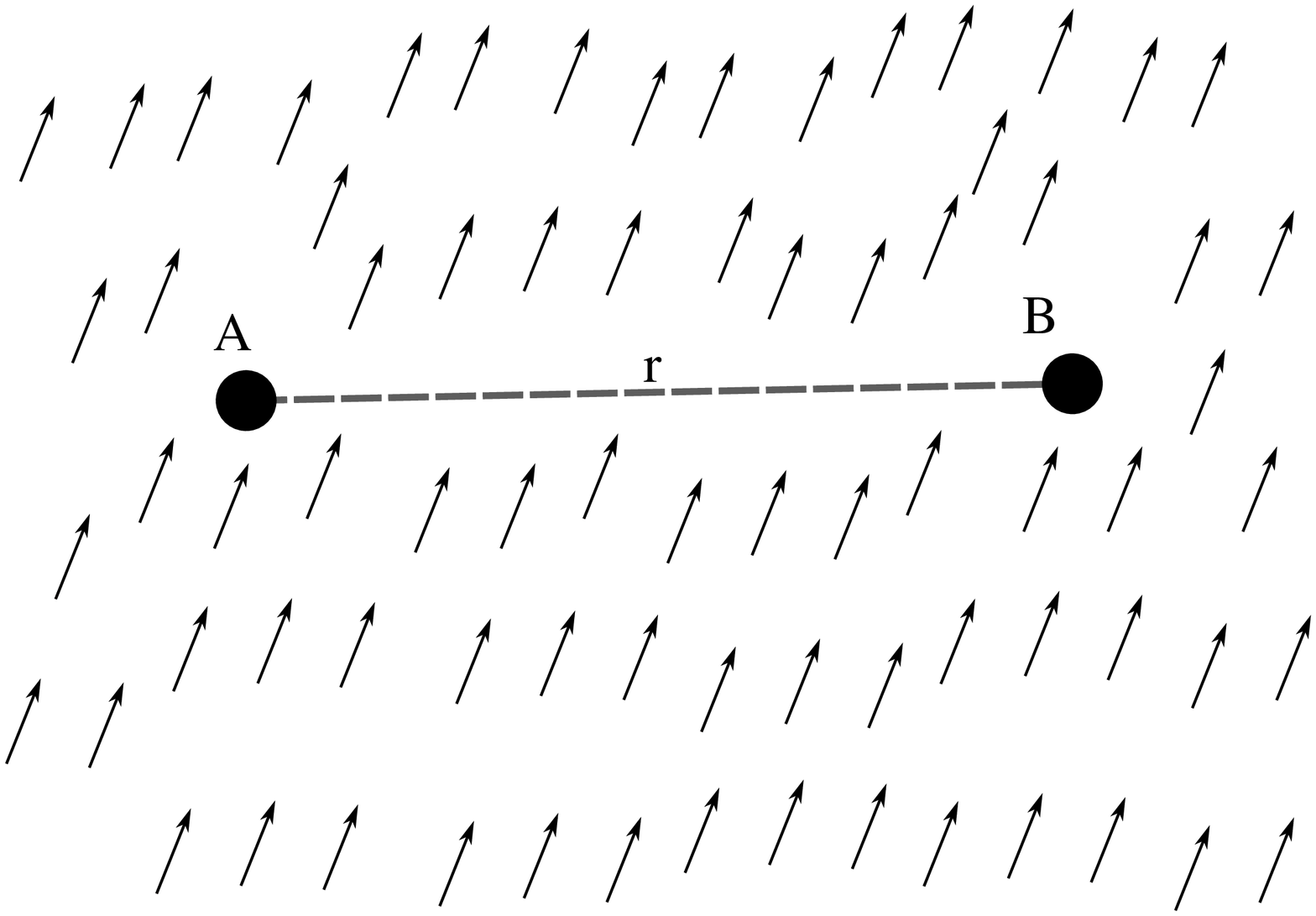}}
$\Bigg\rangle + \Bigg\lvert$
\psfrag{A}{$A$}
\psfrag{B}{$B$}
\psfrag{r}{$r$}
\raisebox{-1cm}{\includegraphics[height=2cm]{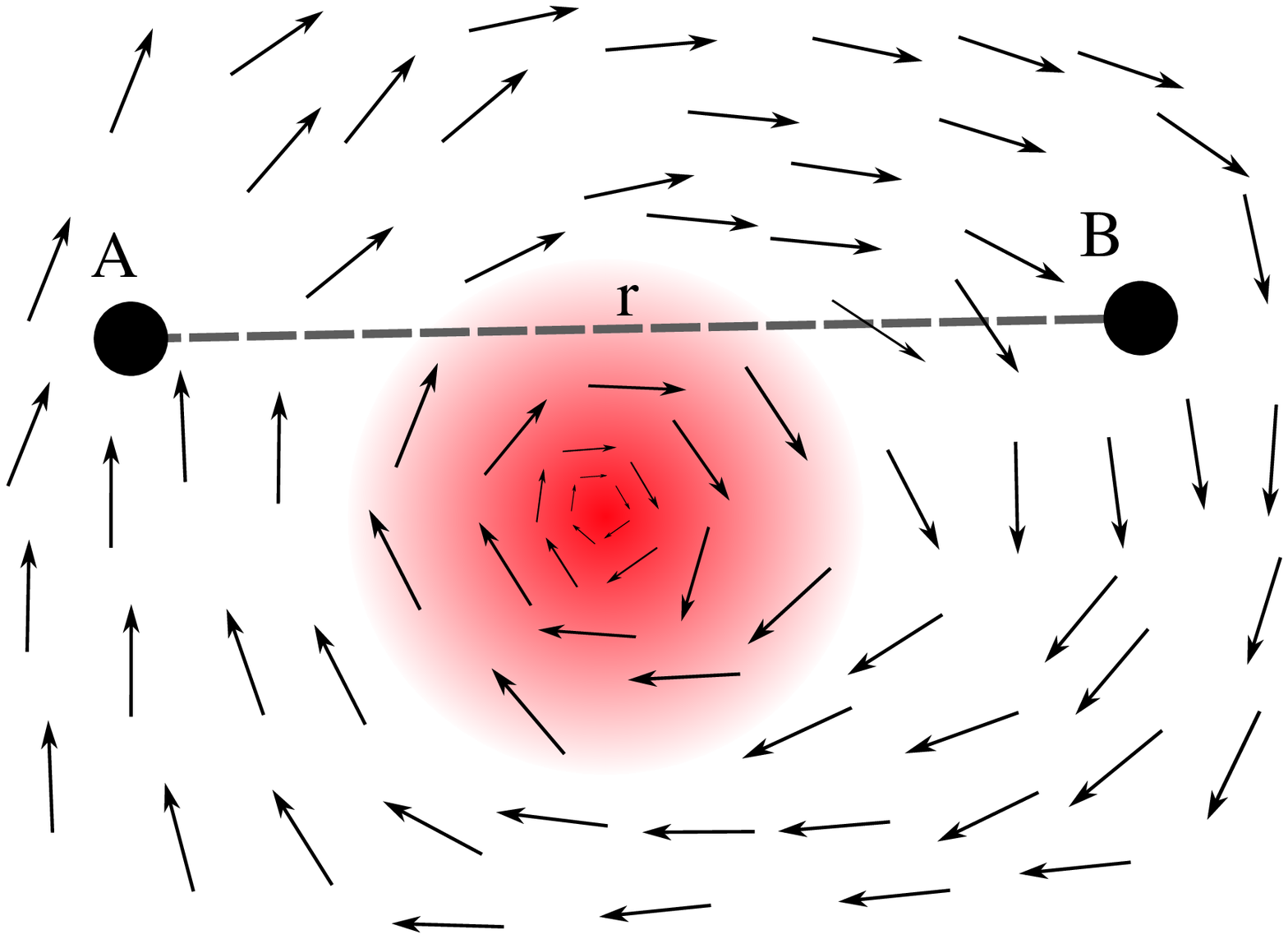}}
$\Bigg\rangle
\Bigg)$
\caption{\label{fig:emergent gauge} When the superfluid phase at patch $A$ is known, the value of the phase at at distant patch $B$ depends on whether or not there is a vortex in between. In the vortex condensate (Higgs phase) vortices can `pop out' of the vacuum spontaneously. The correlations between $A$ and $B$ are in a superposition of `no-vortex' and `vortex' in between. Effectively, the phase at each point can be rotated by an arbitrary amount, i.e. the phase is now emergently gauged.}
\end{center}
\end{figure}

Finally, can we understand the emergent `stay at home' gauge of the Bose Mott-insulator in this dual vortex language?  It is in fact nothing else than the `backward Legendre transformed' version of the demise of the conservation of the supercurrent. This is easy to conceptualize in terms of the effects of vortices on the superfluid order. The Mott scale is just set by the typical distance between free vortex worldlines---at this scale it becomes manifest that sinks and sources are present destroying the supercurrents. Let us now dualize backwards from the currents to the original superfluid phase. Consider the relative orientation of the phase at two patches some length $r$ apart. There might be no vortex in between these two patches such that the phases are corrlated (figure \ref{fig:emergent gauge}). However, when $r$ is larger than the Higgs scale a vortex might occur in the middle, destroying the correlations. In the vortex condensate these possibilities are supposed to occur in coherent superposition and the `no vortex' and `vortex' vacuum configurations are indistinguishable in the same way that a Schr\"odinger cat is as much dead as alive. This implies in turn that the superconducting phase acquires a genuine gauge invariance, the two orientations of the phase at patch $B$ are equally true! 

The take-home message of this section is as follows. The conventional way of deriving the  duality has a `materialistic' attitude, invoking the vortices as a form of matter while the gauge fields enter much in the way as fundamental gauge fields code for the way that  matter interacts. As we discussed, it is however also possible to reformulate the duality in terms of the physical currents, focussing on the way their continuity is lost---in phase representation this turns into the emergent gauge invariance of the Mott insulator.  In the next section we will show that the ingredients of the vortex duality in the gauge language  are strongly dependent on the dimensionality of spacetime, actually posing some problem of principle associated with the nature of string field theory. However, when formulated in terms of the gauge invariant currents  the dependence on dimensionality disappears, just as in the canonical Bose-Hubbard language of section \ref{sec:The Bose-Hubbard model}. This `current language' is still closely tied to the vortex language and this gives us the hold to control the duality in higher dimensions.

\section{The string condensate and duality in the 3+1d $XY$-model}\label{sec:3+1d XY-model}

We have now prepared the reader for the core-section of this paper: how to generalize vortex duality to 3+1 dimensions? In terms of the superfluid phase variables $\varphi(x)$, the story is unchanged: global $U(1)$-symmetry is broken, and there is one massless propagating mode: the spin-wave. Also the correspondence of  the Bose-Mott insulator with the disordered phase (section \ref{sec:The Bose-Hubbard model}) holds. This problem is just equivalent to $XY$ (or $\phi^4$) field theory in 4d---surely a text book problem. But on the dual side things are quite different. The topological defects are now strings tracing out a worldsheet in time (figure \ref{fig:vortex worldsheet}). A worldsheet element is a source $J_{\mu\nu}$ in the sense of Schwinger \cite{Schwinger70}, spanned by two non-parallel spacetime directions, and therefore communicates via the exchange of anti-symmetric 2-form gauge fields $B_{\mu\nu}$.  Let us derive this directly starting from the 3+1d version of the
partition sums  \eqref{eq:generating functional}, \eqref{eq:dual generating functional}. To impose the supercurrent continuity equation $\partial_\mu\xi_\mu =0$ in terms of gauge fields one has to resort to a 2-form Abelian gauge field $B_{\mu \nu}$ \cite{MarshallRamond75,Savit80,Franz07},
\begin{equation}
 \xi_\mu(x) = \epsilon_{\mu\nu\kappa\lambda} \partial_\nu B_{\kappa\lambda}(x).
\end{equation}
The analogue of \eqref{eq:dual gauge action} becomes,
\begin{equation}
 Z_\subscripttext{dual} = \int \mathcal{D} \varphi_\subscripttext{MV} \mathcal{D}B_{\kappa\lambda}\ \te^{\ti \int \frac{1}{2}g (\epsilon_{\mu\nu\kappa\lambda} \partial_\nu B_{\kappa\lambda})^2 + \ti B_{\kappa\lambda} J^V_{\kappa\lambda}}.
\end{equation}

The requirement of the 2-form field to parametrize the continuity equation goes hand in hand with the fact that the vortex is now a worldsheet. The long range vortex--vortex interactions invoke an infinitesimal  worldsheet area, such that the vortex current sourcing the 2-form fields is itself also a 2-form field, 
\begin{equation} 
 J^V_{\kappa\lambda} = \epsilon_{\kappa\lambda \mu \nu} \partial_\mu\partial_\nu \varphi_\subscripttext{MV}.
 \end{equation}
This action is invariant under the gauge transformations
\begin{equation}
 B_{\kappa\lambda} \to B_{\kappa\lambda} + \partial_\kappa \varepsilon_\lambda - \partial_\lambda \varepsilon_\kappa.
\end{equation}
The reader might be less familiar with the counting of the gauge volume of 2-form gauge theories and we have therefore added  \ref{sec:degrees of freedom counting} dealing with these matters  in detail. The bottomline is that of the six independent components of $B_{\kappa\lambda}$, only one is a propagating degree of freedom. This of course corresponds with the `photon' representation of the spin-wave. The 2-form gauge fields are just a fanciful way to take care by extra gauge redundancy that only one propagating mode is associated with the superfluid, instead of the photon doublet that one cannot avoid in a 1-form gauge theory in 3+1d (like electromagnetism).

Obviously, in 3+1d the $XY$-model is no longer dual to electromagnetism as in 2+1d, but instead to a universe of Nielsen--Olesen strings that interact via 2-form gauge fields. In the previous section we learned that the dual formalism also captures the static vortex interactions and in this regard matters are a bit richer in 3+1d. 
Using a coordinate system $(\tau , L , \theta , \phi)$, where $\theta$ and $\phi$ are two orthogonal spatial-transversal directions, and invoking the Coulomb gauge $B_{L\lambda} \equiv 0 \msp \forall \lambda$, the Lagrangian without sources becomes (cf. \eqref{eq:Coulomb gauge}, figure \ref{fig:coordinate systems}),
\begin{equation}\label{eq:3+1d action Coulomb gauge}
 \mathcal{L}_\subscripttext{Coul} = \frac{1}{2}g q^2 B_{\tau \theta}^2 + \frac{1}{2} gq^2 B_{\tau \phi}^2 + \frac{1}{2}g p^2 B_{ \theta \phi}^2.
\end{equation}
The purely transversal component $B_{ \theta \phi}$ is identified as the propagating spin-wave, and the temporal components $B_{\tau \theta}$, $B_{\tau \phi}$ as the static Coulomb forces. The number of Coulomb forces increases because of the higher dimensionality of space: the relative orientation of vortex line sources allows for more diverse interactions. Except for this little surprise, we observe that the  Coulomb phase of  this stringy 2-form gauge theory is coding precisely for the physics of the 3+1d superfluid with its single propagating mode.

Now we want to describe the Higgs phase, the state in which the vortex worldsheet loops grow and extend to the system size. Instead of the worldline tangle of the particle condensate, now a `string condensate' is formed corresponding with a `foam' formed from worldsheets filling spacetime. Currently, there is no way of deriving directly the effective action for such a  Nielsen--Olesen string condensate. This requires knowledge of string field theory, and a second quantized formalism for strings is just not available. Let us recall earlier attempts to generalize the minimal coupling term \eqref{eq:minimal coupling} for stringlike vortices \cite{MarshallRamond75,Rey89,Franz07} (a different path with some ideas similar to ours was taken in \onlinecite{diGreziaEsposito04,diGreziaEspositoNaddeo06}).The defect worldsheet is parametrized by $\sigma = ( \sigma_1 , \sigma_2)$ and $X(\sigma)$ is the map from the worldsheet to real space. Hence each point on the worldsheet $\sigma$ is mapped to a specific point in real space $X(\sigma)$. A surface element of the worldsheet is given by,
\begin{equation}
 \Sigma_{\kappa\lambda}\big[X(\sigma)\big]  = \frac{\partial X_\kappa}{\partial \sigma_1} \frac{\partial X_\lambda}{\partial \sigma_2}  - \frac{\partial X_\lambda}{\partial \sigma_1} \frac{\partial X_\kappa}{\partial \sigma_2}.
\end{equation}
The dynamics of the worldsheet is given by the Nambu--Goto action,
\begin{equation}
 S_\subscripttext{worldsheet} = \int \td^2 \sigma\ T \sqrt{\Sigma_{\mu\nu}\Sigma_{\mu\nu}},
\end{equation}
where the integral is over the entire worldsheet and $T$ is the string tension.

The source term $J_{\kappa\lambda} = \epsilon_{\kappa\lambda \mu \nu} \partial_\mu\partial_\nu \varphi_\subscripttext{MV}$ is related to the worldsheet by,
\begin{equation}
 J_{\kappa\lambda}(x) \sim \int \td^2\sigma\ \Sigma_{\kappa\lambda}\big[X(\sigma)\big]  \delta(X(\sigma) - x).
\end{equation}
According to figure \ref{fig:vortex worldsheet}, the gauge field $B_{\kappa\lambda}(x)$ couples to the worldsheet surface element $\Sigma_{\kappa\lambda}\big[X(\sigma)\big]$. Suppose that a condensate of these vortex strings has formed, giving rise to a collective variable $\Psi\big[X(\sigma)\big]$ which is now a functional of the coordinate function $X(\sigma)$. The fluctuations of the condensate are given by the functional derivative,
\begin{equation}\label{eq:functional derivative}
 \partial_\mu \Psi \to \frac{\delta}{\delta \Sigma_{\kappa\lambda}\big[X(\sigma)\big]} \Psi\big[X(\sigma)\big].
\end{equation}
When a condensate has formed, the amplitude $\lvert \Psi \rvert$ acquires a vacuum expectation value. The amplitude fluctuations freeze as in the particle condensate and only the phase of the string condensate field is left as a dynamical variable. The phase fluctuations enumerate the collective motions of the string condensate but in the absence of an automatic formalism it is guess work to find out what these are. Marshall \& Ramond, Rey and Franz \cite{MarshallRamond75,Rey89,Franz07} find inspiration in the analogy with the particle condensate. The phase degrees of freedom have to be matched through the covariant derivative with the 2-form gauge fields and they conjecture the seemingly obvious generalization,
\begin{equation}\label{eq:vector-valued phase}
  \Psi\big[X(\sigma)\big] = \lvert \Psi \rvert \te^{\ti \int \td X_\mu(\sigma) C_\mu[X(\sigma)]},
\end{equation}
which implies that the collective motions of the string condensate are parametrized in a vector valued phase. The functional derivative \eqref{eq:functional derivative} yields,
\begin{equation}
  \frac{\delta}{\delta \Sigma_{\kappa\lambda}} \Psi\big[X(\sigma)\big] = \lvert \Psi \rvert (\partial_\kappa C_\lambda - \partial_\lambda C_\kappa),
\end{equation}
reducing in turn to a natural minimal coupling form,
\begin{equation}\label{eq:2-form minimal coupling}
 \lvert \frac{\delta}{\delta \Sigma_{\kappa\lambda}} \Psi \rvert \to \lvert (\frac{\delta}{\delta \Sigma_{\kappa\lambda}} - \ti B_{\kappa\lambda}) \Psi \rvert =  \lvert \Psi \rvert (\partial_\kappa C_\lambda - \partial_\lambda C_\kappa - B_{\kappa\lambda}),
\end{equation}
being gauge invariant under the combined transformations,
\begin{align}
 B_{\kappa\lambda} &\to B_{\kappa\lambda} + \partial_\kappa \varepsilon_\lambda - \partial_\lambda \varepsilon_\kappa, \\
C_\kappa &\to C_\kappa + \varepsilon_\kappa.
\end{align}

While this conjecture seems elegant and natural it is actually wrong, at least for the string field theory as of relevance to the 3+1d vortex string condensate. The flaw is in the overcounting of the degrees of freedom of the Mott-insulator/dual superconductor: the vector phase fields ascribes too many collective degrees of freedom to the string condensate. Relying on the gauge invariance in the previous paragraph, we choose the unitary gauge $C_\kappa \equiv 0$ (cf. \eqref{eq:2+1d Higgs action}). The action then reduces to that of a massive 2-form, which is known to have three propagating degrees of freedom. These can be identified by noting that we have `spent' all gauge freedom in this gauge fix, such that all components of $B_{\kappa\lambda}$ become phyiscal degrees of freedom. The three components $B_{\tau \lambda}$ are Coulomb forces, the other three are propagating. But we know that we should end up with two propagating degrees of freedom from the correspondence to the Bose-Mott insulator of section \ref{sec:The Bose-Hubbard model}. Another view on this is that without interactions, this vortex condensate carries the two propagating degrees of freedom of a vector field $C_\kappa$ in four dimensions (just like a photon). In the unitary gauge these two get transferred to the gauge field $B_{\parallel\kappa}$, just as the $\chi$-degree of freedom was transferred to $A_\parallel$ in \eqref{eq:2+1d Higgs action}. So if the vortex condensate were described by \eqref{eq:vector-valued phase}, it would carry two degrees of freedom, instead of only a single pressure mode. 

The absurdity of this guess becomes even more obvious extending matters to higher dimensions. Generalizing this minimal coupling guess to $d$ spacetime dimensions,
\begin{equation}\label{eq:naive minimal couping generalization}
\lvert \partial_\mu \chi - A_\mu \rvert \to \lvert \partial_{[\mu}\chi_{\nu_1 \cdots \nu_{d-3}]} - B_{\mu\nu_1\cdots \nu_{d-3} } \rvert,
\end{equation}
One easy way is to count the number of propagating degrees of freedom of the phase field $\chi_{\nu_1 \cdots \nu_{d-3}}$ if it were not coupled to the gauge field $B_{\mu\nu_1\cdots \nu_{d-3} }$. All of these modes transfer to the gauge field via the Higgs mechanism, adding their degrees of freedom to the single spin-wave mode.
The number of propagating modes for an anti-symmetric form is given by all possible spatial-transversal polarizations (cf. \eqref{eq:3+1d action Coulomb gauge}). In $d$ spacetime dimensions there are $d-2$ transversal directions, which must be accomodated in the $d-3$ indices of the phase field $\chi$. Therefore, the number of degrees of freedom is
\begin{align}
\left( \begin{array}{c} d-2 \\ d-3 \end{array}\right) = \frac{(d-2)!}{(1)!(d-3)!} =  d-2, & &d \ge 3.
\end{align}
This must be added to the single spin-wave mode, so in $d$ spacetime dimensions, the naive prescription \eqref{eq:naive minimal couping generalization} would yield $d-1$ massive degrees of freedom, overcounting the modes of the Mott insulator by $d-3$. In this regard, $d$=2+1 is quite special indeed!

The fact that the usual minimal coupling procedure for the Higgs phenomenon is failing so badly in the higher dimensional cases indicates that it is subtly flawed in a way that does not become obvious in the 2+1d duality case, or even the 3+1d electromagnetic  Higgs condensate. What is then the correct description of the string condensate? It surely has to correspond to the Bose-Mott insulator, which implies that the string condensate 
can only add one extra mode. One way to establish its nature is by invoking a general physics principle: the neutral string condensate would surely represent some form of compressible quantum liquid\label{comment:Soo-Jong Rey}\footnote{It is exactly this point that distinguishes Nielsen--Olesen strings from fundamental strings: the latter are conformally invariant which implies that they \emph{cannot} carry pressure. We thank dr. Soo-Jong Rey for pointing this out. } and such an entity has to carry pressure and thereby a zero sound mode. There is just no room for anything else given the mode counting that we know from the Bose-Mott insulator and we can already conclude that a Nielsen--Olesen string superfluid is at macroscopic distances indistinguishable from a particle superfluid! 

We acquire a full control by employing the gauge invariant current formulation of the duality. The reasoning towards the end of section \ref{sec:2+1d XY-model} pertains as well to the 3+1d case. Regardless the way the currents $\xi_{\mu}$ are parametrized, the `current Higgs action' (\ref{eq:Higgs superflow}) has to be invariably true since it expresses that, due to the fact that the vortex worldlines, strings, whatever destroy the supercurrents, the latter have to acquire mass. In 3+1d one can resolve the non-conserved current fields ($\partial_\mu\xi_\mu \neq 0$) employing  the generalized Helmholtz decomposition \cite{Badur89} for dimensions other than 3. The generalization of \eqref{eq:Helmholtz decomposition} in 3+1d is,
\begin{equation}\label{eq:4d Helmholtz composition}
 \xi_\mu(x) = \partial_\mu \psi(x) + \epsilon_{\mu\nu\kappa\lambda}\partial_\nu B_{\kappa\lambda}(x),
\end{equation}
which holds for any sufficiently smooth four-dimensional vector field that vanishes quicky enough at large distances. As long as current is conserved ($\partial_\mu\xi_\mu = 0$), the first term must be strictly zero. However, we are now dealing with the non-conserved currents and the Helmholtz decomposition demonstrates that this requires the addition of one scalar phase field $\psi$ that takes precisely  the role of the longitudinal photon of the particle condensates---switching off the gauge charge this in turn has to reduce to the zero sound mode of a neutral superfluid. We have now collected all pieces and together with the earlier gauge choices for the static and dynamical gauge fields of the Coulomb phase we can write the effective action of the dual stringy superconductor in 3+1d as,
\begin{align}\label{eq:3+1d Higgs superflow}
 \mathcal{L}_\subscripttext{Higgs}  & =  \frac{1}{2}\xi_\mu(1 + \frac{\Psi_0^2}{g}\frac{1}{-\partial^2}) \xi_\mu \nonumber \\
 & = \frac{1}{2}(p^2 + \frac{\Psi_0^2}{g})( \psi^2 + B_{\perp\theta}^2 + B_{\perp\phi}^2 + B_{\theta\phi}^2).
\end{align}

It is interesting to note that these components of the $B_{\kappa\lambda}$-field are gauge-invariant. In a way, this action is that of Lorenz-gauge-fixed 2-form fields with an additional decoupled scalar field designating the vortex condensate. We identify $\psi$ and $B_{\theta\phi}$ as the two massive propagating degrees of freedom agreeing with the correspondence to the Bose-Mott insulator. The other two terms are the now short-ranged Coulomb forces (cf. \eqref{eq:3+1d action Coulomb gauge}). This leads to the counting scheme laid out in table \ref{table:mode counting}.

\begin{table}
\caption{Mode counting in the $XY$-model}\label{table:mode counting}
\begin{indented}
\item[]\begin{tabular}{c|cc|cc|}
& \multicolumn{2}{|c|}{Coulomb phase} & \multicolumn{2}{|c|}{Higgs phase} \\
&  Coul. forces &  propagating &  Coul. forces &  propagating \\
\hline
 2+1d & 1 long-range& 1 massless& 1 short-range & 2 massive \\
 3+1d & 2 long-range& 1 massless& 2 short-range& 2 massive\\
\hline
\end{tabular}
\end{indented}
\end{table}

\newstuff{This identification of the two propagating modes and two Coulomb forces is based on physical intuition. Is it possible to also capture it within a compact mathematical formulation reflecting the minimal coupling to the condensate field as in \eqref{eq:minimal coupling}? We have argued that it is best to stay in the Lorenz gauge $\partial_\mu B_{\mu\nu}=0$, such that the condensate degree of freedom is represented purely by the phase field $\psi$. The remaining three gauge field components can be collected in a vector field that explicitly removes the longitudinal components that are not physical. To this purpose, one of the indices in the anti-symmetric Levi--Civita tensor is set in the longitudinal direction. This enables us to write down a minimal coupling prescription for two-form fields, analogous to  \eqref{eq:minimal coupling},
\begin{equation}
 \mathcal{L}_\subscripttext{min. coup.}  = \frac{1}{2} \lvert ( \partial_\mu - \ti \epsilon_{\mu \parallel \kappa \lambda}B_{\kappa\lambda}) \Psi \rvert^2.
\end{equation}
When the condensate amplitude is frozen $\lvert \Psi \rvert = \Psi_0$, expansion of this term will lead to the Meissner term in \eqref{eq:3+1d Higgs superflow}.

Thus, through a detour via the physical superflow-variables, we have established the form for minimally coupling a Nielsen--Olesen vortex to a two-form gauge field. The crucial insight is that the longitudinal components of the gauge field are not sourced and should not be taken into consideration. By adding more indices, this form of the minimal coupling can be generalized to even higher dimensions.
}

As we argued, the more precise understanding of the Higgs phenomenon rests on the realization that  the condensate removes the conservation law acting on the fields carrying the forces. The Helmholtz decomposition enumerates precisely the field content. This in turn demonstrates that there is only room for a single scalar longitudinal mode coming from the condensate regardless whether it is formed from particles, strings or the higher-dimensional vortex `branes' encountered in dimensions higher than 3+1d. 
For completeness, we show in \ref{sec:Current conservation in electromagnetism} how to reformulate the Higgs mechanism for standard 3+1d electromagnetism where the heterogeneous Maxwell equation acting on the EM field strength becomes the conservation law being destroyed by the condensate.
\newstuff{
\section{Topological defects in the 3+1d Higgs phase}\label{sec:Topological defects in the 3+1d Higgs phase}
The Higgs phase supports topological defects itself, which we will call Abrikosov vortices even though they communicate via two-form and not vector gauge fields. These vortices are regions where the phase $\chi$ of the collective (superconducting) order parameter field $\Psi$ is singular. By textbook techniques it is readily established that monopole configurations are not stable (like there are no monopoles in a real 3d superconductor), and that the only real topological defects are stringlike. In our dual (gauge-field) language this is quite straightforward, but has a surprising implication: as we have shown above, the Higgs phase of the $XY$ model must correspond to a Bose-Mott insulating state. Therefore the topological excitations in a 3+1d Mott insulator must be stringlike! We will first derive the dynamics of these defects to comment on this interesting point afterwards.

One can repeat the ``dual$^2$'' procedure of \eqref{eq:2+1d dual squared} now for the 3+1d case. We will write down only the most important steps. The minimal coupling term is linearized,
\begin{equation}
 \mathcal{L} = \frac{1}{2} \frac{1}{\Psi_0^2 } v_\mu^2 + \ti v_\mu (\partial_\mu \chi - \epsilon_{\mu \parallel \kappa \lambda}B_{\kappa\lambda}) + \frac{1}{2} g(\epsilon_{\mu\nu\lambda\kappa} \partial_\nu B_{\kappa\lambda})^2.
\end{equation}
The condensate phase $\chi$ is split into a smooth and a multi-valued part. The smooth part is integrated out to give the constraint $\partial_\mu v_\mu = 0$, which is enforced by expressing $v_\mu = \epsilon_{\mu\nu\kappa\lambda} \partial_\nu Z_{\kappa\lambda}$. After several partial integrations and rescaling $B_{\kappa\lambda} \to \frac{1}{\sqrt{g}} B_{\kappa\lambda}$, this leads to,
\begin{align}
\mathcal{L} &= \frac{1}{2} (\epsilon_{\mu\nu\kappa\lambda} \partial_\nu B_{\kappa\lambda})^2 + \frac{1}{2} \frac{1}{\Psi_0^2} (\epsilon_{\mu\nu\kappa\lambda} \partial_\nu Z_{\kappa\lambda})^2 \nonumber\\
& \phantom{=} + \ti  Z_{\kappa\lambda} K_{\kappa\lambda} - \ti\frac{1}{\sqrt{g}}  Z_{\kappa\lambda} \epsilon_{\kappa\lambda\mu\nu} \partial_\nu \epsilon_{\mu\parallel \rho \sigma} B_{\rho \sigma},
\end{align}
where $K_{\kappa\lambda} = \epsilon_{\kappa\lambda\mu\nu}\partial_\mu\partial_\nu \chi_\subscripttext{MV}$ is the Abrikosov vortex current. From this form one sees that the Abrikosov vortices are stringlike, since $K_{\kappa\lambda}(x)$ describes a surface element of the vortex worldsheet (cf. figure  \ref{fig:vortex worldsheet}). For contractions in the last term we use the identity 
\begin{equation}
 \epsilon_{\kappa\lambda\mu\parallel}  \epsilon_{\mu\parallel \rho \sigma} = \delta_{\kappa\rho}\delta_{\lambda \sigma} - \delta_{\kappa\sigma}\delta_{\lambda \rho},
\end{equation}
where the indices on the right-hand side take values orthogonal to $\parallel$ only. The coupling of the $Z$-gauge field to the $B$-gauge field then looks like,
\begin{equation}
 \ti\frac{1}{\sqrt{g}}  Z_{\kappa\lambda} \epsilon_{\kappa\lambda\parallel\mu}  (\epsilon_{\mu\nu\rho \sigma} \partial_\nu B_{\rho \sigma})  = \ti\frac{1}{\sqrt{g}}  Z_{\kappa\lambda} \epsilon_{\kappa\lambda\parallel\mu} \xi_\mu.
\end{equation}

The gauge field $B_{\rho\sigma}$ only shows up in the combination $\xi_\mu = \epsilon_{\mu\nu\rho \sigma} \partial_\nu B_{\rho \sigma}$, which can be integrated out to yield a Meissner term for $Z_{\kappa\lambda}$,
\begin{equation}
\mathcal{L} =  \frac{1}{2} \frac{1}{\Psi_0^2} (\epsilon_{\mu\nu\kappa\lambda} \partial_\nu Z_{\kappa\lambda})^2 + \frac{1}{2g}  Z_{\kappa\lambda}^2 + \ti  Z_{\kappa\lambda} K_{\kappa\lambda} ,
\end{equation}
which is valid in the Lorenz gauge $\partial_\kappa Z_{\kappa\lambda}  = 0$. Here we have a theory of Abrikosov vortex strings $K_{\kappa\lambda}$ that have short-range interactions with each other through the exchange of massive two-form fields $Z_{\kappa\lambda}$. When vortices proliferate, they are described by a collective field $\Phi$, minimally coupled to the gauge field that we have rescaled $Z_{\kappa\lambda} \to \Psi_0 Z_{\kappa\lambda}$,
\begin{align}
 \mathcal{L} &= \frac{1}{2} (\epsilon_{\mu\nu\kappa\lambda} \partial_\nu Z_{\kappa\lambda})^2 + \frac{\Psi_0^2}{2g}Z_{\kappa\lambda}^2 \nonumber\\
&\phantom{=} + \frac{1}{2}\lvert(\partial_\mu - \ti \Psi_0 \epsilon_{\mu\parallel\kappa\lambda} Z_{\kappa\lambda})\Phi\rvert^2 + \frac{1}{2} M^2 \lvert\Phi\rvert^2 + \frac{1}{4} W \lvert \Phi \rvert^4.
\end{align}
Through the phase transition, the Abrikosov vortices destroy the `superconducting' order so that $\Psi_0$ vanishes. Then the gauge field $Z_{\kappa\lambda}$ decouples and we are left with the action of a neutral superfluid \eqref{eq:neutral superfluid}, exactly our starting point. In this way dual$^2 = 1$ also holds in 3+1 dimensions.

Now we return to the interpretation of these results. In the Meissner phase there are vortex solutions with a finite core size, that cause (dual) supercurrent to flow around them within a shell of thickness inversely proportional to the Higgs mass $\Psi_0^2 /g$. This thickness is called the \emph{penetration depth}. In a real superconductor the vortices are caused by an external magnetic field $\mathbf{B} = \nabla \wedge \mathbf{A}$, and it is the vector potential that sources the supercurrent. In our case the Meissner phase is equivalent to the Bose-Mott insulator. The equivalent of the magnetic field is the spatial curl of the gauge field which is given by the temporal component of the superfluid current,
\begin{equation}
 \xi_\tau = \cases{ \epsilon_{\tau i j } \partial_i A_j & \text{2+1d},\\
             \epsilon_{\tau i j k} \partial_i B_{jk} & \text{3+1d}.
            }
\end{equation}
From this, we conclude that defects in the 3+1d Bose-Mott insulator are stringlike regions where superfluid order persists locally. It is the converse to the statement that vortices in the superfluid are regions where dual superconducting order $\Psi$ persists. We can therefore crudely think up the following experiment: one would create,  perhaps in a cold atoms on an optical lattice setup \cite{GreinerEtAl02,BruderFazioSchoen05}, a slab of Mott insulating state sandwiched between regions of superfluid order (figure \ref{fig:vortex lines in a Bose-Mott insulator}). The Mott insulator must be tuned to exactly integer filling. For the correct values of other parameters involved, the proximity coherence length of superfluid order may be so large that it can penetrate into the Mott insulator. This would then demonstrate the existence of vortex lines in the Bose-Mott insulator.
\begin{figure}
 \begin{center}
\psfrag{s}{superfluid}
\includegraphics[height=2.4cm]{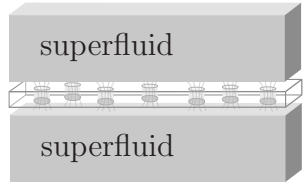}
\caption{\label{fig:vortex lines in a Bose-Mott insulator} Proposed setup to show vortex lines in the Bose-Mott insulator. \small{The Mott insulator (white) should be sandwiched between two regions with superfluid order (grey). The order parameter extends outside of the superfluid itself to pierce through the Mott insulator, in the form of vortex lines.}}
\end{center}
\end{figure}
This would be a surprise: up till now the common knowledge was that the Bose-Mott insulator supports particle-like excitations in the form of the doublon and holon modes. In the dual language those are represented by the gauge fields. But the sources of those modes turn out to be topological excitations which are $p$-branes. The physics of the Bose-Mott insulator is therefore richer than previously expected.

It is important to realize that this behaviour mimics that of type-II superconductors. There an external magnetic flux can penetrate at field strengths much lower than the naively expected critical field, since it need not destroy superconduction order completely, but penetrate only in small regions, the Abrikosov vortices. Similarly, if one could create a bias in superfluid density $\sim \xi_\tau$ across the Mott insulating slab, supercurrent will flow through thin wires. This would be a sort of ``type-II Josephson current''.
}

\section{Conclusions}\label{sec:Conclusions}

Intrigued by the fact that the dualities associated with the most primitive field theories ($XY$/$\phi^4$/Bose-Hubbard in 3+1- and higher dimensions) are not textbook material we focussed in on the 3+1d case. A simple string field theory problem (the vortex worldsheet foam) lies at the heart of this lacking knowledge. Resting on the detailed understanding of the disordered state in terms of the Bose-Mott insulator that corresponds with this Nielsen--Olesen `string superconductor' we precisely named the field theory describing its effective properties. In fact, we were forced to abandon the standard minimal coupling construction of the Higgs phenomenon that confused earlier attempts to construct the dual theory. The `longitudinal photons' of the standard Higgs mechanism are in fact obscuring constructions and the misleading nature becomes obvious at the moment one generalizes away from the particle condensates. Within the confines of the dual superconductors associated with superfluids, we emphasized that the dual Higgs mechanism is essentially rooted in the demise of the supercurrents of the superfluid. The vortex condensate destroys their conservation and through the Helmholtz construction one learns that in the effective theory of the dual superconductor there is only room for one extra scalar longitudinal mode. Via this 
detour we learn that the condensate formed from Nielsen--Olesen strings is quite dull: the only collective mode it sustains is zero sound, and in this regard it is at long distances indistiguishable from the standard particle Bose-condensate!

In fact, there is nothing special to the 3+1d case and we arrive at the main conclusion in this paper:  {\em the neutral superfluid--charged superconductor duality of the 2+1d global $U(1)$ theory is equally valid in $D+1$ dimensional systems with $D>2$, where the dual superconductor describes a $D-1$ form gauge theory Higgsed by a $p=D-2$ Nielsen--Olesen brane condensate that supports one massive compressional mode.}

 It might already be obvious to the reader but let us finish this exposition by an explicit derivation of this statement: 

For each broken symmetry generator, there is a Goldstone mode that communicates the ridigity of that order parameter. The set of Goldstone modes $\{ \varphi^a \}$ is labelled by an index $a$. Because these modes are massless and non-interacting, the canonical momenta $\xi^a_\mu = \frac{\partial \mathcal{L}}{\partial (\partial_\mu \varphi^a)}$ are conserved $\partial_\mu \xi_\mu^a = 0$. They are in fact the Noether currents under the global symmetry transformations $\varphi^a(x) \to \varphi^a(x) + \alpha^a$. As current carries energy, the action is of the form $S \sim \int \xi_\mu^a\xi_\mu^a$.
Topological defects are regions where the Goldstone variable is not well-defined; consequently, the current is no longer conserved in that region. Each flavour $a$ of current $\xi^a_\mu$ can be generated by the appropriate topological defect. A condensate of such defects $\Psi^a$ will have two effects: i) they generate current everywhere, so that it is conserved nowhere $\partial_\mu \xi^a_\mu \neq 0$ which introduces a new degree of freedom; ii) the current--current correlations are destroyed by the defects, causing them to be exponentially decay with scale set by the Higgs mass $\Psi^a_0$. The action in the Higgs phase is of the form,
\begin{equation}
 S \sim \int \xi_\mu^a\big(1 + \frac{(\Psi_0^a)^2}{-\partial^2}\big)\xi_\mu^a.
\end{equation}
Each current has a description in terms of anti-symmetric $d-2$-form gauge fields ($d= D+1$),
\begin{equation}
 \xi^a_\mu = \partial_\mu \psi^a + \epsilon_{\mu\nu\lambda_1 \cdots \lambda_{d-2} }\partial_\nu B^a_{\lambda_1 \cdots \lambda_{d-2} }.
\end{equation}
The longitudinal components $B^a_{\parallel \lambda_2 \cdots \lambda_{d-2}}$ are unphysical and can be gauged away. The components $B^a_{\perp \lambda_2 \cdots \lambda_{d-2}}$ correspond to the $d-2$ Coulomb forces per $a$ between the $d-3$-brane defects. For each $a$ there is one more component corresponding to the propgating Goldstone mode. The scalar fields $\psi^a$ vanish in the Coulomb phase and are dynamic condensate modes in the Higgs phase. They may be represented by the symmetric purely longitudinal components $B^a_{\parallel \ldots \parallel}$.

\newstuff{
Another result is that the dual gauge formalism directly identifies the topological defects of the Bose-Mott insulator, which are particle-like in 2+1d but string-like in 3+1d. These vortices may be induced by nearby superfluid order. This is a bit of a surprise and shows the power of duality construction. We have given a crude idea of how to find these vortices in an experimental setup.
}

It would be interesting to see how well this scheme holds for other actual physical systems. The related case of Abelian--Higgs model or scalar QED or Ginzburg--Landau theory in 3+1d is treated in \ref{sec:Current conservation in electromagnetism}. One interesting suitable problem should be the physics of (quantum) liquid crystals \cite{CvetkovicZaanen06b}, in which the interplay between rotational and translational defects complicates matters.

\ack{We thank Koenraad Schalm and Jian-Huang She for useful discussions and dr. Soo-Jong Rey for the comment mentioned on p.\pageref{comment:Soo-Jong Rey}. This work was supported by the Netherlands foundation for Fundamental Research of Matter (FOM) and the Nederlandse Organisatie voor Wetenschappelijk Onderzoek (NWO) via a Spinoza grant.}

\appendix
\section{Degrees of freedom counting}\label{sec:degrees of freedom counting}
We have determined the degrees of freedom by explicit examination of the action and propagators. There is a more general and formal way of deriving the \emph{propagating} degrees of freedom given an action (Coulomb forces do not fall into this general scheme). It precisely determines the gauge degrees of freedom and the influence of constraints. This is exhaustively explained in Ref \onlinecite{HenneauxTeitelboim92}. We will very briefly discuss this procedure for free Abelian 1- and 2-forms (\onlinecite{HenneauxTeitelboim92} ch.19).

The Maxwell Lagrangian in $d$ spacetime dimensions is,
\begin{equation}\label{eq:Maxwell Lagrangian}
 \mathcal{L} = -\frac{1}{4} F_{\mu\nu}^2 = - \frac{1}{2} ( \partial_\mu A_\nu - \partial_\nu A_\mu)^2.
\end{equation}
The vector field $A_\mu$ has $d$ components, so we start out with $d$ degrees of freedom. The action is invariant under gauge transformation $A_\mu \to A_\mu + \partial_\mu \varepsilon$; furthermore this gauge transformation corresponds to a so-called \emph{first-class constraint}, which means it removes two degrees of freedom in total. The reason for this is that we fix the vector field not only in space at one moment in time (a time slice), but also its evolution using $\partial_\tau \varepsilon$. Another point of view is that the temporal component $A_\tau$ is set by the scalar electrostatic potential, which is zero everywhere for a free field; the temporal component is completely fixed by the equation of motion $\nabla^2 A_\tau = 0$.

Therefore a free vector field in $d$ dimensions has $d-2$ propagating degrees of freedom, exactly the transversal polarizations of the photon.

The generalization of \eqref{eq:Maxwell Lagrangian} for an anti-symmetric 2-form field $B_{\mu\nu}$ in 4 dimensions is,
\begin{equation}
 \mathcal{L} = -\frac{1}{2} (\epsilon_{\mu\nu\kappa\lambda} \partial_\nu B_{\kappa\lambda})^2.
\end{equation}
The field has six independent components. The action is invariant under gauge transformations,
\begin{equation}\label{eq:2-form gauge transformation}
 B_{\kappa\lambda}(x) \to B_{\kappa\lambda}(x) + \partial_\kappa \varepsilon_\lambda(x)  -  \partial_\lambda \varepsilon_\kappa(x).
\end{equation}
Here $\varepsilon_\lambda(x)$ is any smooth real vector field with 4 components; but there are only three independent gauge transformations since $\delta_{\lambda\kappa}( \partial_\kappa \varepsilon_\lambda -\partial_\lambda \varepsilon_\kappa ) =0$ always. As explained above each gauge transformation removes two degrees of freedom. The transformations are however redundant, since another vector field,
\begin{equation}
 \varepsilon'_\lambda(x) = \varepsilon_\lambda(x) + \partial_\lambda \eta(x),
\end{equation}
where $\eta$ is any smooth scalar field gives exactly the same transformation in \eqref{eq:2-form gauge transformation}. A free 2-form field in 4 dimensions therefore has $6-(6-1) = 1$ propagating degree of freedom.

\section{Current conservation in electromagnetism}\label{sec:Current conservation in electromagnetism}
We apply the conservation-of-current considerations to the most famous example of the Higgs mechanism: the photon field in 3+1 dimensions coupled to a complex scalar condensate field. This is variously known as the Abelian--Higgs model, Ginzburg--Landau theory or scalar QED. It describes the basic physics of the electromagnetic field in the vacuum and in a superconductor.

The electromagnetic field is a vector field $A_\mu(x)$. Its dynamics is gouverned by the field strength $F_{\mu\nu} = \partial_\mu A_\mu - \partial_\nu A_\mu$ and the Maxwell action,
\begin{equation}
 S = \int  -\frac{1}{4} F^2_{\mu\nu}.
\end{equation}
The field strength is invariant under the gauge transformation $A_\mu \to A_\mu + \partial_\mu \varepsilon$. The vector field with gauge fix $\partial_\mu A_\mu =0$ has three degrees of freedom: the two transversal photon polarizations $A_\theta$ and $A_\phi$, and the part mediating static Coulomb interations $A_\perp$. 

The field strength $F_{\mu\nu}$ has six independent components and is therefore overcounting the degrees of freedom. This can be cured by imposing the homogeneous Maxwell equations,
\begin{equation}
 \td \mathsf{F} = \epsilon_{\mu\nu\kappa\lambda} \partial_\nu F_{\kappa\lambda} = 0.
\end{equation}
In $(\parallel,\perp,\theta ,\phi)$-coordinates (see figure \ref{fig:coordinate systems}) this implies that the only non-zero components of the field strength are $F_{\parallel \nu}$, which we collect in a vector field $f_\nu \equiv F_{\parallel \nu}$ (the `current'). From this point we act as if the field strength $F_{\parallel \nu}$ is not necessarily anti-symmetric; still the longitudinal component is set to zero as long as there are no external sources: $\partial_\nu f_\nu = \partial_\nu F_{\parallel \nu} = J^\text{ext}_\parallel \to 0$ (inhomogeneous Maxwell equations). The other three components of $f_\nu$ correspond to the three physical degrees of freedom identified above via,
\begin{equation}\label{eq:physical electromagnetic current}
 f_\nu = p A_\nu.
\end{equation}
Now we couple the photon field to a complex scalar Higgs field via $\lvert \partial_\mu \Psi \rvert \to \lvert (\partial_\mu -\ti A_\mu)\Psi \rvert$ as in \eqref{eq:minimal coupling}. The Higgs field describes a condensate destroying the current conservation, so that the longitudinal component $f_\parallel$ is released. Indeed, from \eqref{eq:physical electromagnetic current} this corresponds to the longitudinal polarization of the photon: $f_\parallel = p A_\parallel$. In terms of the field strength, it is seen to correspond to the symmetric component $F_{\parallel,\parallel}$, which is normally not taken into consideration. 
\section*{References}
\bibliography{references}
\end{document}